\documentclass[letters,fleqn,usenatbib,useAMS]{mnras}

\usepackage{mathptmx}
\usepackage[T1]{fontenc}
\usepackage{ae,aecompl}

\usepackage{graphicx}	% Including figure files
\usepackage{amsmath}	% Advanced maths commands
\usepackage{amssymb}	% Extra maths symbols

\newcommand{\kms}{km\,s$^{-1}$}

\title[The pulsating Bp star HD\,96446]{
The challenge of measuring magnetic fields in strongly pulsating stars: the case of HD\,96446
}

\author[S. P. J\"arvinen et al.]{
S.~P.~J\"arvinen,$^{1}$\thanks{E-mail: sjarvinen@aip.de}
S.~Hubrig,$^{1}$,
I.~Ilyin$^{1}$,
M.~Sch\"oller$^{2}$, and
M.~Briquet$^{1,3}$
\\
$^{1}$Leibniz-Institut f\"ur Astrophysik Potsdam (AIP), An der Sternwarte~16, D-14482~Potsdam, Germany\\
$^{2}$European Southern Observatory, Karl-Schwarzschild-Str.~2, D-85748~Garching, Germany\\
$^{3}$Institut d'Astrophysique et de G\'eophysique, Universit\'e de Li\`ege, Quartier Agora, All\'ee du 6 ao\^ut 19C, B-4000 Li\`ege, Belgium
}

\date{Accepted 2016 September 16. Received 2016 Sepbember 16; in
  original form 2016 July 19}

\pubyear{2016}

\begin{document}
\label{firstpage}
\pagerange{\pageref{firstpage}--\pageref{lastpage}}
\maketitle

\begin{abstract}
  Among the early B-type stars, He-rich Bp stars exhibit the strongest
  large-scale organized magnetic fields with a predominant dipole
  contribution. The presence of $\beta$\,Cep-like pulsations in the typical
  magnetic early Bp-type star HD\,96446 was announced a few years ago, but
  the analysis of the magnetic field geometry was hampered by the absence of
  a reliable rotation period and a sophisticated procedure for accounting for
  the impact of pulsations on the magnetic field measurements. Using new
  spectropolarimetric observations and a recently determined rotation period
  based on an extensive spectroscopic time series, we investigate the magnetic
  field model parameters of this star under more detailed considerations of
  the pulsation behaviour of the line profiles.
\end{abstract}


%%%%%%%%%%%%%%%%%%%%%%%%%%%%%%%%%%%%%%%%%%%%%%%%%%

%%%%%%%%%%%%%%%%% BODY OF PAPER %%%%%%%%%%%%%%%%%%

\section{Introduction}
\label{sec:intro}

HD\,96446 is a typical magnetic He-strong B2p star exhibiting spectral
variations of the He and Si lines. This star was extensively discussed by 
\citet{neiner2012} 
who detected $\beta$\,Cep-like radial velocity variations with a period of
about 2.23\,h and employed all available magnetic field measurements to
estimate magnetic model parameters. During the last years, more
spectropolarimetric material for this star became available. In addition, a
rotation period of 23.4\,d was recently identified using an extensive
spectroscopical time series data set obtained by 
Briquet et al.\ (in preparation).
Here, we re-discuss the magnetic field model parameters using a specific field
measurement strategy to account for pulsational variability both in radial
velocity shifts and in the profile shapes.

%----------------------------------------------------------------------------

\section{Observations and data reduction}
\label{sec:obs}

We obtained spectropolarimetric observations of HD\,96446 with the
High Accuracy Radial velocity Planet Searcher polarimeter
\citep[HARPSpol;][]{snik2008} 
attached to ESO's 3.6\,m telescope (La Silla, Chile) on 2016 June 14 and 15.
Each observation was split into four subexposures with (sub-)exposure times
varying between 23 and 30\,min. The quarter-wave retarder plate was rotated
by $90\degr$ after each subexposure. Further, we downloaded older HARPSpol
observations from the ESO archive that were obtained on the nights of 2011
May 23, 24, and 27, as well as on 2012 July 14--19 and 22. The four HARPSpol
observations obtained on 2011 May were already presented in the paper of 
\citet{neiner2012}.
All archival data have (sub-)exposure times of 15\,min. A summary of all
HARPSpol observations is given in Table~\ref{tab:log}. The columns give the
heliocentric Julian date (HJD) for the middle of the subexposures and the
signal-to-noise ratio (S/N) of the spectra. All spectra have a resolving power
of about $R = 115\,000$ and cover the spectral range 3780--6910\,\AA{}, with
a small gap between 5259 and 5337\,\AA{}. The reduction and calibration of
these spectra was performed using the HARPS data reduction software available
at the ESO headquarter in Germany and on La~Silla. The normalization of the
spectra to the continuum level is described in detail by 
\citet{Hubrig2013}.

%---------------------- T1
\begin{table}
\centering
\caption{
Logbook of the all individual subexposures of the spectropolarimetric
observations of HD\,96446 with HARPSpol.
}
\label{tab:log}
\begin{tabular}{crp{3mm}cr}
\hline\hline
\multicolumn{1}{c}{HJD} &
\multicolumn{1}{c}{S/N} &
&
\multicolumn{1}{c}{HJD} &
\multicolumn{1}{c}{S/N} \\
\hline
2455704.5157 & 72  & & 2456125.4683 & 115 \\
2455704.5265 & 75  & & 2456125.4790 & 111 \\
2455704.5373 & 90  & & 2456125.4899 & 125 \\
2455704.5481 & 92  & & 2456125.5007 & 120 \\
2455705.6247 & 129 & & 2456126.4599 & 141 \\
2455705.6355 & 126 & & 2456126.4707 & 144 \\
2455705.6463 & 111 & & 2456126.4815 & 148 \\
2455705.6571 & 107 & & 2456126.4923 & 153 \\
2455706.4470 & 192 & & 2456128.4634 & 210 \\
2455706.4578 & 185 & & 2456128.4742 & 197 \\
2455706.4686 & 212 & & 2456128.4850 & 182 \\
2455706.4794 & 249 & & 2456128.4958 & 200 \\
2455709.4335 & 169 & & 2456131.4560 & 142 \\
2455709.4443 & 133 & & 2456131.4668 & 143 \\
2455709.4551 & 149 & & 2456131.4776 & 138 \\
2455709.4659 & 121 & & 2456131.4884 & 137 \\
2456123.4555 & 82  & & 2457554.4462 & 193 \\
2456123.4664 & 75  & & 2457554.4634 & 246 \\
2456123.4772 & 99  & & 2457554.4811 & 274 \\
2456123.4880 & 87  & & 2457554.4982 & 339 \\
2456124.4482 & 177 & & 2457555.5065 & 260 \\
2456124.4590 & 172 & & 2457555.5277 & 277 \\
2456124.4698 & 154 & & 2457555.5478 & 300 \\
2456124.4806 & 163 & & 2457555.5667 & 280 \\
\hline
\end{tabular}
\end{table}

%----------------------------------------------------------------------------

\section{Pulsations and magnetic field analysis}
\label{sec:mfield}

\begin{figure}
 \centering
 	\includegraphics[width=\columnwidth]{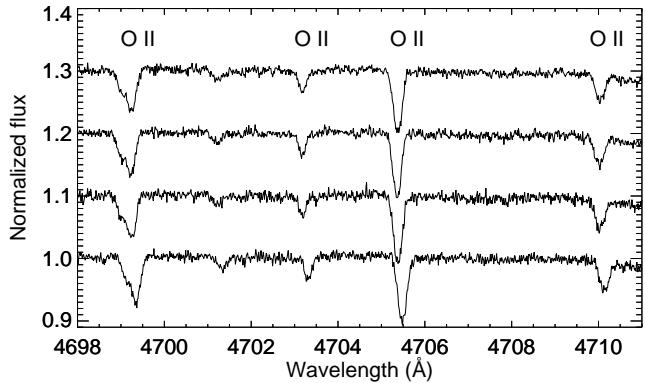}
       \caption{
         Changes in line profile shapes between subexposures shown for the
         third 2011 data set.
         }
   \label{fig:lpch}
\end{figure}

As already discussed by 
\citet{neiner2012},
the null spectra ($N$), usually used to diagnose spurious polarization
signatures, of HD\,96446 do not appear flat. Their spectral analysis revealed
shifts in radial velocity between subexposures due to pulsations, and they
concluded that the signatures in the null spectra are directly linked to
pulsational radial velocity shifts. To exclude the impact of pulsations on the
magnetic field measurements, before combining all subexposures, the authors
corrected individual spectra by their measured radial velocity. Since
$\beta$\,Cep pulsational line profile variability, in addition to radial
velocity shifts, also has an impact on the line profile shape
\citep[see fig.~3 of][]{hubrig2011},
we decided to carry out magnetic field measurements between the
left-hand-polarized and right-hand-polarized spectra for each subexposure
separately ($(I \pm V)_{\alpha}$ with $\alpha = 45^{\circ}, 135^{\circ}, 225^{\circ}, 315^{\circ}$). The final longitudinal magnetic field value for each epoch was
then calculated as an average of the measurements for each subexposure. An
example of line profile changes in each subexposure obtained at four different
angles is presented in Fig.~\ref{fig:lpch}.

%---------------------- F1

\begin{figure*}
 \centering
 	\includegraphics[width=\textwidth]{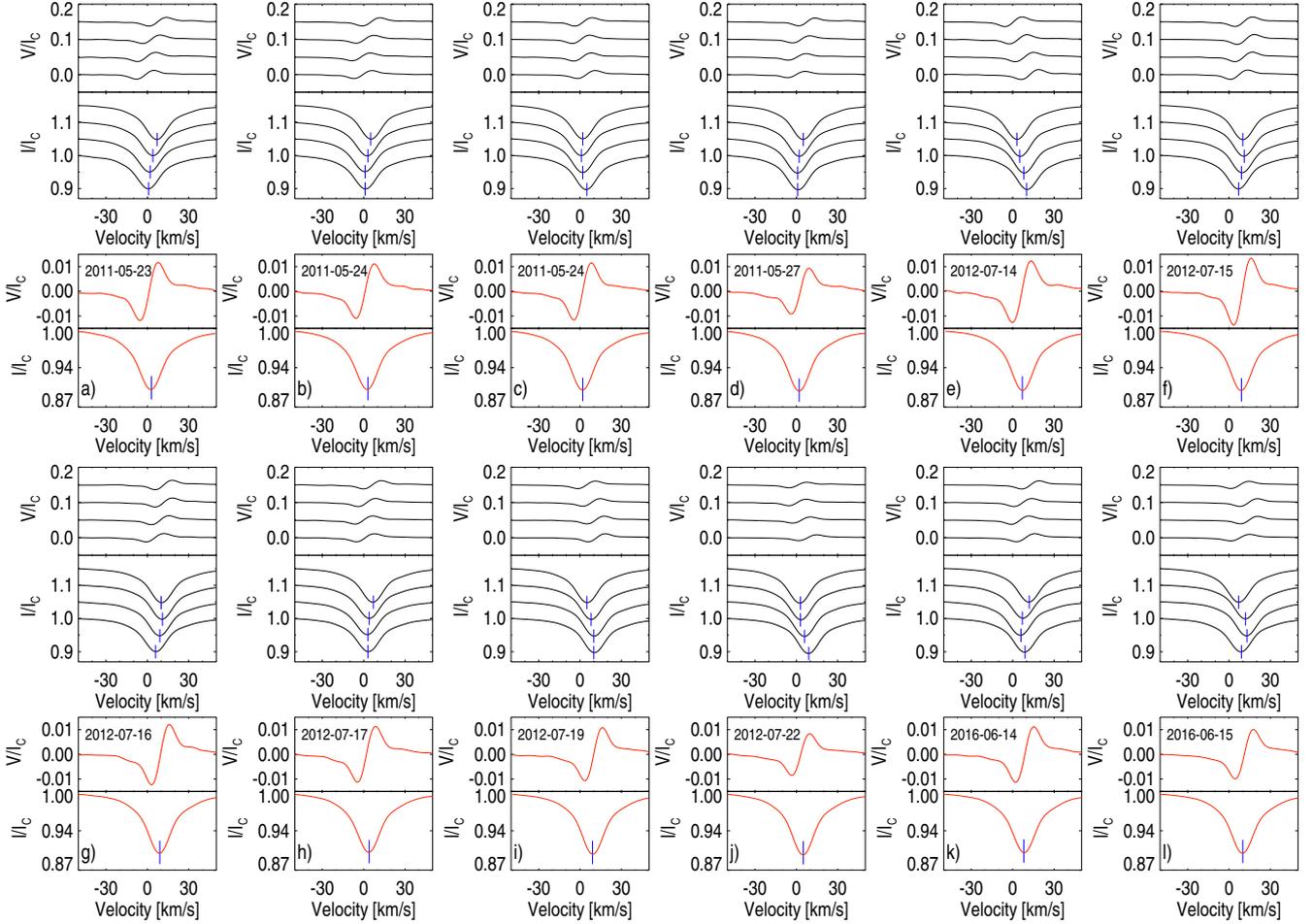}
       \caption{
         Stokes~$V$ (on top) and $I$ (on bottom) LSD profiles for individual
         subexposures (upper panels) and corresponding final Stokes~$V$ and
         $I$ profiles after having moved the subexposures into the same
         reference velocity frame (lower panels) for 12 different epochs
         (a--l). The blue ticks in the Stokes~$I$ profiles indicate line
         centres.
       }
   \label{fig:profiles}
\end{figure*}

\begin{figure}
 \centering
 \includegraphics[width=\columnwidth]{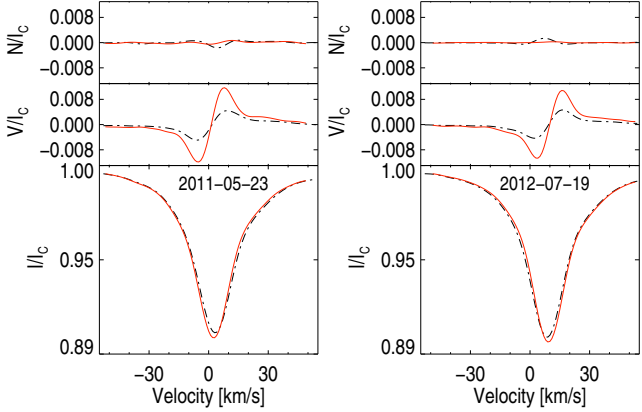}
 \caption{
   Two examples of Stokes~$I$, $V$, and $N$ profile differences for cases where velocity shifts
   between subexposures have been ignored (black dash-dot line) and where
   they have been corrected for (red solid line). These plots correspond to subplots (a) and (i) in
   Fig.~\ref{fig:profiles}. The profiles have been calculated using all
   lines.
 }
   \label{fig:profcomp}
\end{figure}

Given the rather low S/N of the individual subexposures, we employed the
least-squares deconvolution (LSD) technique, allowing us to achieve much
higher S/N in the LSD spectra. The details of this technique and of the
calculation of  Stokes $I$ and $V$ parameters can be found in the work of
\citet{Donati1997}.
A line mask containing 196 metallic (C, N, O, Mg, Al, Si, S, and Fe) and He
lines is constructed using the Vienna Atomic Line Database
\citep[VALD; e.g.][]{kupka2000,vald2}
based on the stellar parameters obtained for HD\,96446 by
\citet{neiner2012}.
The longitudinal magnetic field $\left<B_{\rm z}\right>$ for each subexposure
is measured by calculating the first-order moment of the Stokes $V$ profile
\citep[e.g.][]{Mathys1989}:
\begin{equation}
\left<B_{\mathrm z}\right> = -2.4 \times 10^{11}\frac{\int \upsilon V
  (\upsilon){\mathrm d}\upsilon }{\lambda_{0}g_{0}c\int
  [I_{c}-I(\upsilon )]{\mathrm d}\upsilon},
\end{equation}
where $\upsilon$ is the Doppler velocity in \,km\,s$^{-1}$, and
$\lambda_{0}$ and $g_{0}$ are the normalization values of the wavelength and
the average Land\'e factor. The adopted velocity range was $\pm20$~\kms around
the line centre. To illustrate the velocity shifts in the LSD profiles
obtained using this line list between the four subexposures, we present them
in the upper panels of Fig.~\ref{fig:profiles} for 12 epochs and for both
Stokes~$V$ (top) and $I$ (bottom). In the lower panels we show the Stokes~$V$
(top) and $I$ (bottom) profiles calculated from subexposures after the
velocity shifts have been removed. Fig.~\ref{fig:profcomp} shows typical
differences between velocity shift uncorrected and corrected profiles.
Obviously, ignoring the velocity shifts leads to an underestimation of the
magnetic field strength. While the velocity corrected profile of the first
example on Fig.~\ref{fig:profcomp} gives $-1189\pm68$\,G, the uncorrected
profile leads only to $-582\pm43$\,G, and the corresponding values for the
second example are $-997\pm60$\,G versus $-499\pm47$\,G. In
Table~\ref{tab:MF}, we present our measurements of the longitudinal magnetic
field $\left< B_{\rm z}\right>$. Column~1 gives the HJD, Column~2 the rotation
phase according to the rotation period of 23.38\,d (see below), Column~3 the
S/N of the LSD spectrum obtained for `All', and the next three columns refer
to the $\left< B_{\rm z}\right>$ measurements for three different line masks.
`All' has 196 lines with a mean Land\'e factor $\bar{g}_{\rm eff} = 1.21$,
`He' has 20 lines with $\bar{g}_{\rm eff} = 1.25$,
and `Si' has 14 lines with $\bar{g}_{\rm eff} = 1.18$.
Column~7 presents the results from 
\citet{neiner2012} 
using only metal lines.

%---------------------- F2

%---------------------- F3

%---------------------- T2
\begin{table*}
\centering
\caption{Longitudinal magnetic field values of HD\,96446 computed from
 HARPSpol data.
}
\label{tab:MF}
\begin{tabular}{crcr@{$\pm$}lr@{$\pm$}lr@{$\pm$}lr@{$\pm$}l}
\hline\hline
\multicolumn{1}{c}{HJD} &
\multicolumn{1}{c}{Phase} &
\multicolumn{1}{c}{S/N} &
\multicolumn{2}{c}{$\left<B_{\rm z}\right>_{\rm All}$} &
\multicolumn{2}{c}{$\left<B_{\rm z}\right>_{\rm He}$} &
\multicolumn{2}{c}{$\left<B_{\rm z}\right>_{\rm Si}$} &
\multicolumn{2}{c}{$\left<B_{\rm z}\right>_{\rm metal}$} \\
&
&
&
\multicolumn{2}{c}{(G)} &
\multicolumn{2}{c}{(G)} &
\multicolumn{2}{c}{(G)} &
\multicolumn{2}{c}{(G)} \\
\hline\\
2455704.540 & 0.13 & 1133 & $-$1189 & 68  & $-$1295 & 144 & $-$2108 & 256 & $-$1075 & 38 \\
2455705.649 & 0.18 & 1271 & $-$1129 & 39  & $-$1216 & 45  & $-$2006 & 100 & $-$960  & 29 \\
2455706.471 & 0.22 & 1713 & $-$1113 & 24  & $-$1224 & 44  & $-$2056 & 80  & $-$932  & 17 \\
2455709.457 & 0.34 & 1360 & $-$896  & 38  & $-$1167 & 38  & $-$1585 & 141 & $-$738  & 21 \\
2456123.477 & 0.05 & 1267 & $-$1216 & 46  & $-$1261 & 46  & $-$2213 & 179 \\
2456124.470 & 0.10 & 1615 & $-$1198 & 21  & $-$1247 & 62  & $-$2247 & 178 \\
2456125.490 & 0.14 & 1300 & $-$1142 & 46  & $-$1217 & 113 & $-$2081 & 71  \\
2456126.481 & 0.18 & 1606 & $-$1143 & 24  & $-$1223 & 28  & $-$2015 & 144 \\
2456128.485 & 0.27 & 1653 & $-$997  & 60  & $-$1193 & 23  & $-$2034 & 70  \\
2456131.477 & 0.40 & 1309 & $-$823  & 37  & $-$1217 & 10  & $-$1584 & 192 \\
2457554.472 & 0.26 & 1851 & $-$1093 & 47  & $-$1224 & 29  & $-$2033 & 100 \\
2457555.537 & 0.30 & 1871 & $-$953  & 51  & $-$1211 & 27  & $-$1794 & 143 \\
\hline
\end{tabular}
\end{table*}

\begin{figure*}
 \centering
 	\includegraphics[width=\textwidth]{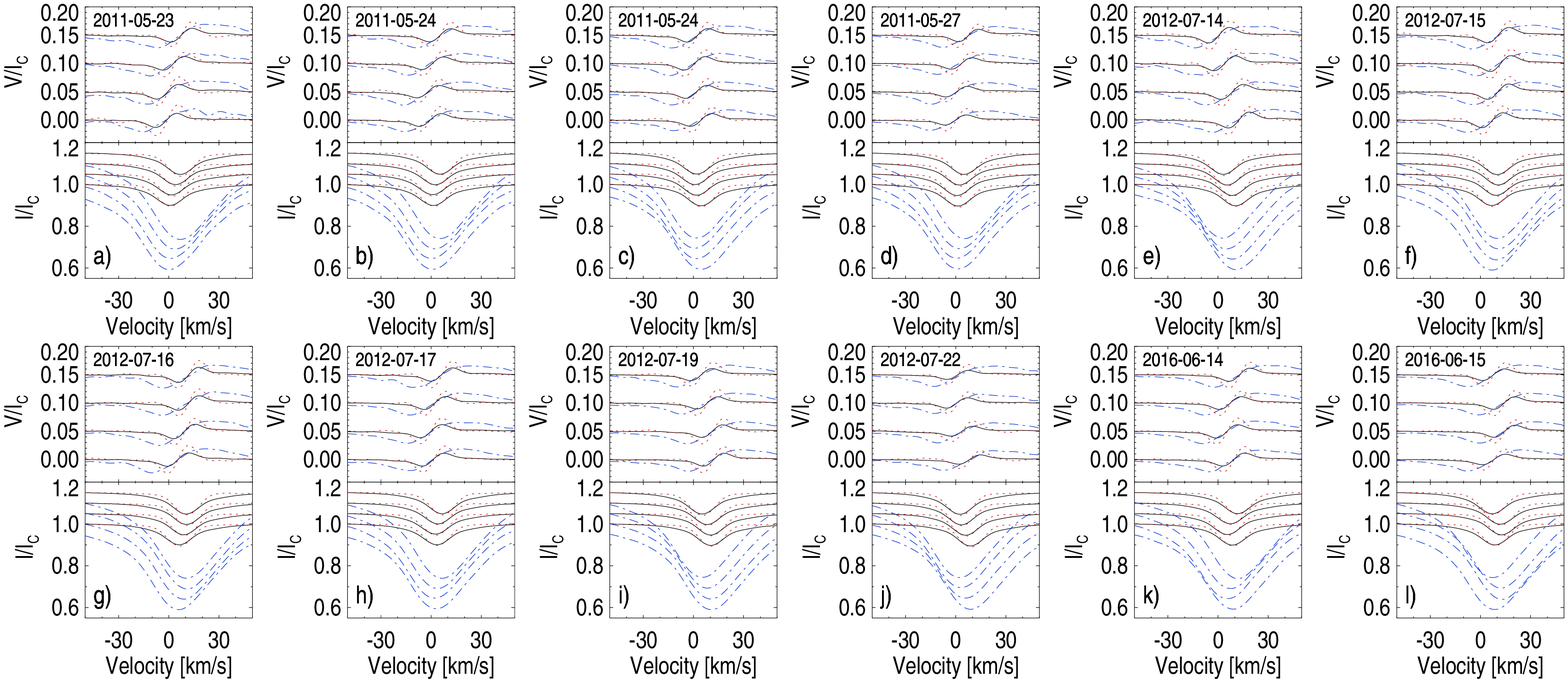}
       \caption{
         Comparison of Stokes~$V$ (top) and $I$ (bottom) LSD profiles for
         different line masks. The black solid lines are as in
         Fig.~\ref{fig:profiles} using all 196 lines, the blue dash-dotted
         lines represent the He line profiles (20 lines), and the red dotted
         lines are for Si (14 lines).
       }
   \label{fig:elemprofs}
\end{figure*}

Since He-rich Bp stars usually exhibit surface elemental chemical
inhomogeneities, best detectable from the variability of line profiles
belonging to the elements He and Si, we also studied the magnetic field
strength distribution measured using the samples of He and Si lines
separately. A comparison of the Stokes~$I$ and $V$ LSD profiles for different
samples is presented in Fig.~\ref{fig:elemprofs}. Magnetic field strength
values can be found in the literature, i.e., from
\citet{borra1979} and
\citet{bohlender1987},
who both used H$\beta$ lines, and from
\citet{Mathys1994},
whose results are based on selected metallic lines.
Similarly, the values obtained by
\citet{neiner2012}
are obtained from metal lines and correspond to our first four measurements
using the same spectra (see Table~\ref{tab:MF}). The obtained values are in
good agreement, taking into account that we have some He lines in our line
list.

%---------------------- F4

Due to the presence of surface chemical inhomogeneities in Bp stars, the
rotation period of the star can be determined by examining the equivalent
width variability of several spectral lines. Based on an extensive time
series of high-resolution spectra for HD\,96446,
Briquet et al. (in preparation) 
obtained a rotation period of 23.4\,d.

Under the assumption of a dipolar magnetic field, the longitudinal magnetic
field measurements $\left< B_{\rm z}\right>$ vary as a sine wave. Given our
HARPSpol observations over a long time-base, we can refine the value of the
rotation frequency $f_{\rm rot}$, performing a non-linear least-squares
sine-wave fit to the $\left< B_{\rm z}\right>$-values for all three line lists.
From this, we get $f_{\rm rot} = 0.04277$ d$^{-1}$ for `All'. The values
obtained when using the magnetic data for the Si lines and for the He lines
are fully compatible with the frequency deduced for all lines within
0.00005 d$^{-1}$. The corresponding rotation period is
$P_{\rm rot} = 23.38 \pm 0.0273$\,d.

\begin{figure}
 \centering
 \includegraphics[width=\columnwidth]{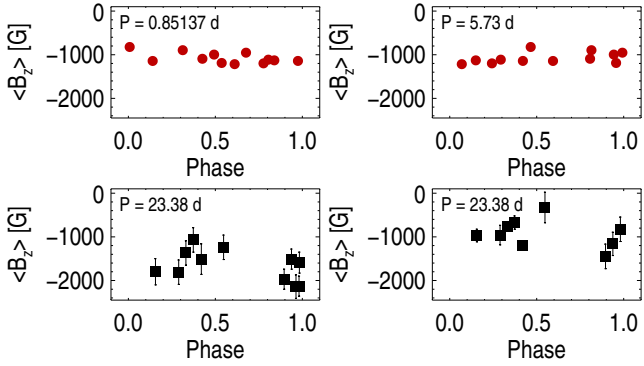}
 \caption{
   Magnetic field measurement distribution over the rotational phase. On the
   top panels, red filled circles represent our measurement values that are
   phased using the two period values as considered in literature and in the
   lower panels the phases for older literature H$\beta$ (left) and
   metal (right) values are based on our refined rotation period.
 }
   \label{fig:pha}
\end{figure}

\begin{figure}
 \centering
 \includegraphics[width=\columnwidth]{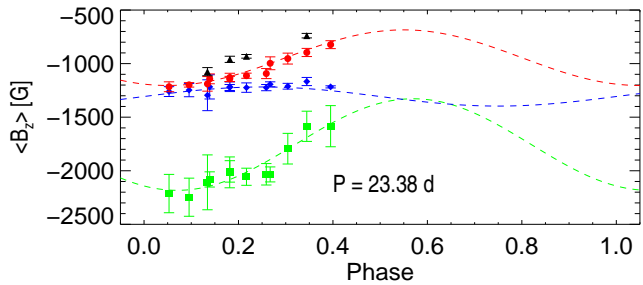}
 \caption{
   Magnetic field measurement distribution over the rotational phase using
   the ephemeris $T = 2\,455\,000.0 + 23\fd38 \ E$. Red filled circles
   represent values measured from all lines, green squares are from Si lines,
   and blue diamonds are from He lines. Dashed lines represent the fits. In
   addition, the values measured by
   \citet{neiner2012}
   using only metal lines are denoted with black triangles. 
 }
   \label{fig:epha}
\end{figure}

In the upper panel of Fig.~\ref{fig:pha}, we present our magnetic field measurements
phased with a period of 0.85137\,d as suggested by 
\citet{matthewsbohlender} 
and a period of 5.73\,d as suggested by 
\citet{neiner2012}.
In the two lower panels, we present the measurements taken from the
literature phased with a period of 23.38\,d, separately for the
measurements using H$\beta$ and for those using metal lines. The
corresponding phase diagrams for the mean longitudinal magnetic field
measurements using three different line masks, including the best sinusoidal
fits, are presented in Fig.~\ref{fig:epha} where we also show the measurements
by
\citet{neiner2012}
that were based on selected metallic lines. The increasing trend in field
strength between phases 0.1 and 0.4 is also present in the older measurements
that were calculated using H$\beta$ (see Fig.~\ref{fig:pha}, lower left-hand
panel). The difference between the measurements carried out for the line lists
containing He and Si separately shows that the corresponding field strengths
are differing at some rotation phases by about 1\,kG (see also Columns~5 and
6 in Table~\ref{tab:MF}). The measurements based on the He lines show an
almost flat distribution.

 %---------------------- F5

%---------------------- F6

Assuming that HD\,96446 exhibits a single-wave variation of the longitudinal
magnetic field during the stellar rotation cycle, we can estimate the
parameters of the dipole configuration. Using the oblique rotator model 
\citep{stibbs},
$v \sin i = 3\pm 2$\,km\,s$^{-1}$ and $R = 4.45 \pm 2.0\,R\odot$ as
given by 
\citet{neiner2012},
and our newly refined rotation period  $P_{\rm rot} = 23.38 \pm 0.0273$\,d,
we obtain an inclination angle of the stellar rotation axis to the line of
sight $i=18\pm15^{\circ}$ and the equatorial velocity 
$v_{\rm eq}=10\pm4$\,km\,s$^{-1}$. The variation of the phase curve for the
measurements using all lines has a mean of 
$\overline{\left< B_{\rm z}\right>} = -945\pm 26$\,G and an amplitude of
$A_{\left< B_{\rm z}\right>} = 260 \pm 19$\,G. From this, we obtain
$\left< B_{\rm z} \right>^{\rm min}=-685\pm32$\,G and
$\left< B_{\rm z} \right>^{\rm max}=-1205\pm32$\,G.
Using the definition by 
\citet{Preston1967}

\begin{equation}
r = \frac{\left< B_{\rm z}\right>^{\rm min}}{\left< B_{\rm z}\right>^{\rm max}}
  = \frac{\cos \beta \cos i - \sin \beta \sin i}{\cos \beta \cos i + \sin \beta
\sin i},
\end{equation}

\noindent
we find $r=0.568\pm0.026$ and finally following

\begin{equation}
\beta =  \arctan \left[ \left( \frac{1-r}{1+r} \right) \cot i \right],
\label{eqn:4}
\end{equation}

\noindent
we obtain the magnetic obliquity angle $\beta=40\pm25^{\circ}$. 
Assuming the limb-darkening coefficient $u = 0.28\pm0.02$  based on the
calculations of
\citet{2011A&A...529A..75C}
for visual wavelengths and atmospheric parameters of HD\,96446 we estimate
the dipole strength to be $4.64\pm0.88$\,kG.

\begin{figure}
  \centering
  \includegraphics[width=0.8\columnwidth]{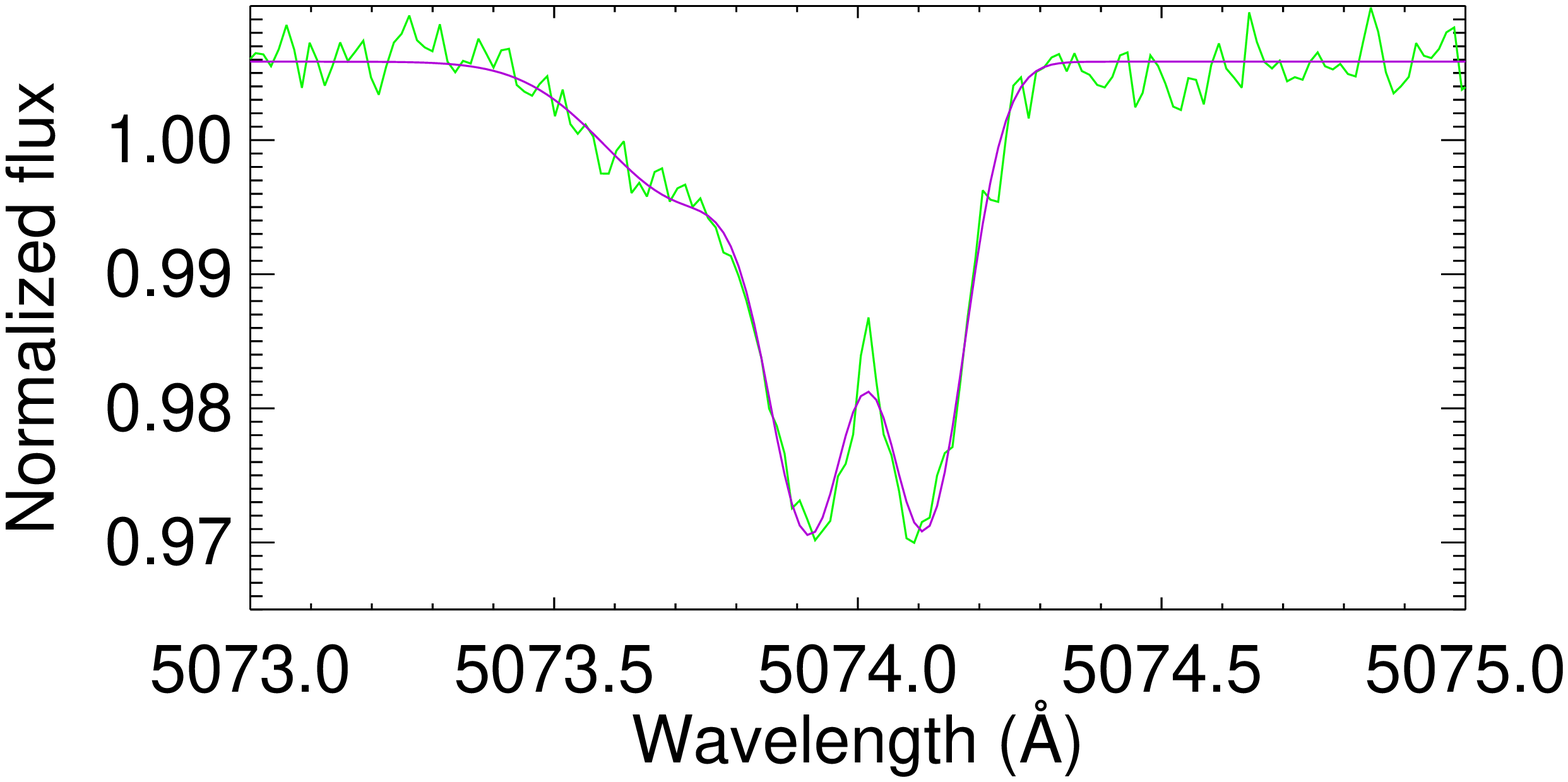}
  \caption{
    The mean profile of the 2016 June 14 subexposures showing the
    magnetically split \ion{Fe}{iii} line at 5073.9~\AA{} (green) and a
    triple Gaussian fit to it (purple) to take into account the blend in the
    blue wing caused by the \ion{N}{ii} line at 5073.6~\AA{}.
  }
    \label{fig:split}
\end{figure}

As already mentioned in the work of 
\citet{neiner2012}, 
the magnetic splitting is best visible in the \ion{Fe}{III} line at 
5073.9~\AA{} (Fig.~\ref{fig:split}). From this magnetically split line,
it is possible to determine the mean magnetic field modulus, i.e.\ the
average over the visible stellar hemisphere of the modulus of the
magnetic field vector that is weighted by the local line intensity,
following the relation given, e.g.\ by
\citet{hubrig2007}.
The observations from 2016 June 14 ($\varphi=0.26$) lead to
$\langle|B|\rangle=3.9\pm0.25$~kG.

%---------------------- F7

%----------------------------------------------------------------------------

\section{Conclusions}

For the analysis of the magnetic field behaviour in the star HD\,96446 over
the refined rotation period of 23.38\,d, we used a sample of archival
HARPSpol observations completed by our own most recent observations taken at
much higher S/N with the same instrument. We show the strong impact of
pulsations on the line profiles and the magnetic field measurements and
suggest a specific measurement strategy to account for pulsational
variability. For the first time the measurements are carried out for the
elements He and Si separately and show that the individual field strengths
are differing at some rotation phases by about 1\,kG. These elements are
usually distributed inhomogeneously over the stellar surface of Bp stars.
Model parameters for HD\,96446 were constrained assuming a simple magnetic
dipole. Compared to the model parameters presented by 
\citet{neiner2012}, 
the inclination angle $i$ deduced in our analysis appears to be slightly
larger while the value for the magnetic obliquity is in a similar range. While 
\citet{neiner2012}
estimate the dipole strength between 5 and 10\,kG, our modelling results in
a much lower value of $4.64\pm0.88$\,kG, which is also consistent with the
estimation of the magnetic field modulus of $3.9\pm0.25$\,kG. Due to the
relative brightness of HD\,96446, the presence of a rather strong magnetic
field, and its slow rotation,  it can be considered as an excellent target for
future high-resolution spectropolarimetric observations over the whole
rotation cycle to improve our understanding of the magnetic field structure in
massive early B-type stars and the relation between the magnetic structure
and the distribution of surface chemical inhomogeneities.

\section*{Acknowledgements}

Based on observations made with ESO Telescopes at the La Silla Paranal
Observatory under programme IDs 187.D-0917(A,C) (PI: Alecian) and
097.C-0277(A) (PI: Hubrig).

This work has made use of the VALD database, operated at Uppsala
University, the Institute of Astronomy RAS in Moscow, and the University of Vienna.

SPJ acknowledges support by the Deutsche Forschungsgemeinschaft, grant
JA~2499/1--1.

%%%%%%%%%%%%%%%%%%%%%%%%%%%%%%%%%%%%%%%%%%%%%%%%%%



\bibliographystyle{mnras}
\bibliography{hd96446} % if your bibtex file is called example.bib

\begin{thebibliography}{}
\makeatletter
\relax
\def\mn@urlcharsother{\let\do\@makeother \do\$\do\&\do\#\do\^\do\_\do\%\do\~}
\def\mn@doi{\begingroup\mn@urlcharsother \@ifnextchar [ {\mn@doi@}
  {\mn@doi@[]}}
\def\mn@doi@[#1]#2{\def\@tempa{#1}\ifx\@tempa\@empty \href
  {http://dx.doi.org/#2} {doi:#2}\else \href {http://dx.doi.org/#2} {#1}\fi
  \endgroup}
\def\mn@eprint#1#2{\mn@eprint@#1:#2::\@nil}
\def\mn@eprint@arXiv#1{\href {http://arxiv.org/abs/#1} {{\tt arXiv:#1}}}
\def\mn@eprint@dblp#1{\href {http://dblp.uni-trier.de/rec/bibtex/#1.xml}
  {dblp:#1}}
\def\mn@eprint@#1:#2:#3:#4\@nil{\def\@tempa {#1}\def\@tempb {#2}\def\@tempc
  {#3}\ifx \@tempc \@empty \let \@tempc \@tempb \let \@tempb \@tempa \fi \ifx
  \@tempb \@empty \def\@tempb {arXiv}\fi \@ifundefined
  {mn@eprint@\@tempb}{\@tempb:\@tempc}{\expandafter \expandafter \csname
  mn@eprint@\@tempb\endcsname \expandafter{\@tempc}}}

\bibitem[\protect\citeauthoryear{{Bohlender}, {Landstreet}, {Brown}  \&
  {Thompson}}{{Bohlender} et~al.}{1987}]{bohlender1987}
{Bohlender} D.~A.,  {Landstreet} J.~D.,  {Brown} D.~N.,   {Thompson} I.~B.,
  1987, \mn@doi [\apj] {10.1086/165830}, \href
  {http://adsabs.harvard.edu/abs/1987ApJ...323..325B} {323, 325}

\bibitem[\protect\citeauthoryear{{Borra} \& {Landstreet}}{{Borra} \&
  {Landstreet}}{1979}]{borra1979}
{Borra} E.~F.,  {Landstreet} J.~D.,  1979, \mn@doi [\apj] {10.1086/156907},
  \href {http://adsabs.harvard.edu/abs/1979ApJ...228..809B} {228, 809}

\bibitem[\protect\citeauthoryear{{Claret} \& {Bloemen}}{{Claret} \&
  {Bloemen}}{2011}]{2011A&A...529A..75C}
{Claret} A.,  {Bloemen} S.,  2011, \mn@doi [\aap]
  {10.1051/0004-6361/201116451}, \href
  {http://cdsads.u-strasbg.fr/abs/2011A%26A...529A..75C} {529, A75}

\bibitem[\protect\citeauthoryear{{Donati}, {Semel}, {Carter}, {Rees}  \&
  {Collier Cameron}}{{Donati} et~al.}{1997}]{Donati1997}
{Donati} J.-F.,  {Semel} M.,  {Carter} B.~D.,  {Rees} D.~E.,   {Collier
  Cameron} A.,  1997, \mnras, \href
  {http://cdsads.u-strasbg.fr/abs/1997MNRAS.291..658D} {291, 658}

\bibitem[\protect\citeauthoryear{{Hubrig} \& {Nesvacil}}{{Hubrig} \&
  {Nesvacil}}{2007}]{hubrig2007}
{Hubrig} S.,  {Nesvacil} N.,  2007, \mn@doi [\mnras]
  {10.1111/j.1745-3933.2007.00310.x}, \href
  {http://cdsads.u-strasbg.fr/abs/2007MNRAS.378L..16H} {378, L16}

\bibitem[\protect\citeauthoryear{{Hubrig}, {Ilyin}, {Briquet}, {Sch{\"o}ller},
  {Gonz{\'a}lez}, {Nu{\~n}ez}, {De Cat}  \& {Morel}}{{Hubrig}
  et~al.}{2011}]{hubrig2011}
{Hubrig} S.,  {Ilyin} I.,  {Briquet} M.,  {Sch{\"o}ller} M.,  {Gonz{\'a}lez}
  J.~F.,  {Nu{\~n}ez} N.,  {De Cat} P.,   {Morel} T.,  2011, \mn@doi [\aap]
  {10.1051/0004-6361/201117261}, \href
  {http://cdsads.u-strasbg.fr/abs/2011A%26A...531L..20H} {531, L20}

\bibitem[\protect\citeauthoryear{{Hubrig}, {Ilyin}, {Sch{\"o}ller}  \& {Lo
  Curto}}{{Hubrig} et~al.}{2013}]{Hubrig2013}
{Hubrig} S.,  {Ilyin} I.,  {Sch{\"o}ller} M.,   {Lo Curto} G.,  2013, \mn@doi
  [Astron. Nachr.] {10.1002/asna.201311948}, \href
  {http://cdsads.u-strasbg.fr/abs/2013AN....334.1093H} {334, 1093}

\bibitem[\protect\citeauthoryear{{Kupka}, {Ryabchikova}, {Piskunov}, {Stempels}
   \& {Weiss}}{{Kupka} et~al.}{2000}]{kupka2000}
{Kupka} F.~G.,  {Ryabchikova} T.~A.,  {Piskunov} N.~E.,  {Stempels} H.~C.,
  {Weiss} W.~W.,  2000, Balt.~Astron., \href
  {http://cdsads.u-strasbg.fr/abs/2000BaltA...9..590K} {9, 590}

\bibitem[\protect\citeauthoryear{{Mathys}}{{Mathys}}{1989}]{Mathys1989}
{Mathys} G.,  1989, Fundam.~Cosm.~Phys., \href
  {http://esoads.eso.org/abs/1989FCPh...13..143M} {13, 143}

\bibitem[\protect\citeauthoryear{{Mathys}}{{Mathys}}{1994}]{Mathys1994}
{Mathys} G.,  1994, \aaps, \href
  {http://cdsads.u-strasbg.fr/abs/1994A%26AS..108..547M} {108, 547}

\bibitem[\protect\citeauthoryear{{Matthews} \& {Bohlender}}{{Matthews} \&
  {Bohlender}}{1991}]{matthewsbohlender}
{Matthews} J.~M.,  {Bohlender} D.~A.,  1991, \aap, \href
  {http://cdsads.u-strasbg.fr/abs/1991A%26A...243..148M} {243, 148}

\bibitem[\protect\citeauthoryear{{Neiner}, {Landstreet}, {Alecian}, {Owocki},
  {Kochukhov}, {Bohlender}  \& {MiMeS Collaboration}}{{Neiner}
  et~al.}{2012}]{neiner2012}
{Neiner} C.,  {Landstreet} J.~D.,  {Alecian} E.,  {Owocki} S.,  {Kochukhov} O.,
   {Bohlender} D.,   {MiMeS Collaboration} 2012, \mn@doi [\aap]
  {10.1051/0004-6361/201218988}, \href
  {http://adsabs.harvard.edu/abs/2012A%26A...546A..44N} {546, A44}

\bibitem[\protect\citeauthoryear{{Preston}}{{Preston}}{1967}]{Preston1967}
{Preston} G.~W.,  1967, \mn@doi [\apj] {10.1086/149358}, \href
  {http://cdsads.u-strasbg.fr/abs/1967ApJ...150..547P} {150, 547}

\bibitem[\protect\citeauthoryear{{Ryabchikova}, {Piskunov}, {Kurucz},
  {Stempels}, {Heiter}, {Pakhomov}  \& {Barklem}}{{Ryabchikova}
  et~al.}{2015}]{vald2}
{Ryabchikova} T.,  {Piskunov} N.,  {Kurucz} R.~L.,  {Stempels} H.~C.,  {Heiter}
  U.,  {Pakhomov} Y.,   {Barklem} P.~S.,  2015, \mn@doi [\physscr]
  {10.1088/0031-8949/90/5/054005}, \href
  {http://adsabs.harvard.edu/abs/2015PhyS...90e4005R} {90, 054005}

\bibitem[\protect\citeauthoryear{{Snik}, {Jeffers}, {Keller}, {Piskunov},
  {Kochukhov}, {Valenti}  \& {Johns-Krull}}{{Snik} et~al.}{2008}]{snik2008}
{Snik} F.,  {Jeffers} S.,  {Keller} C.,  {Piskunov} N.,  {Kochukhov} O.,
  {Valenti} J.,   {Johns-Krull} C.,  2008, in SPIE Conf. Series. p.~22,
  \mn@doi{10.1117/12.787393}

\bibitem[\protect\citeauthoryear{{Stibbs}}{{Stibbs}}{1950}]{stibbs}
{Stibbs} D.~W.~N.,  1950, \mn@doi [\mnras] {10.1093/mnras/110.4.395}, \href
  {http://cdsads.u-strasbg.fr/abs/1950MNRAS.110..395S} {110, 395}

\makeatother
\end{thebibliography}

%%%%%%%%%%%%%%%%%%%%%%%%%%%%%%%%%%%%%%%%%%%%%%%%%%

%%%%%%%%%%%%%%%%%%%%%%%%%%%%%%%%%%%%%%%%%%%%%%%%%%

% Don't change these lines
\bsp	% typesetting comment
\label{lastpage}
\end{document}